\documentclass[final,5p,times,twocolumn]{elsarticle}

\usepackage{amssymb}
\usepackage{amsmath}
\usepackage{xcolor}
\usepackage{siunitx}
\begin{document}
	
	\begin{frontmatter}
		
		%% Title, authors and addresses
		
		%% use the tnoteref command within \title for footnotes;
		%% use the tnotetext command for theassociated footnote;
		%% use the fnref command within \author or \address for footnotes;
		%% use the fntext command for theassociated footnote;
		%% use the corref command within \author for corresponding author footnotes;
		%% use the cortext command for theassociated footnote;
		%% use the ead command for the email address,
		%% and the form \ead[url] for the home page:
		%% \title{Title\tnoteref{label1}}
		%% \tnotetext[label1]{}
		%% \author{Name\corref{cor1}\fnref{label2}}
		%% \ead{email address}
		%% \ead[url]{home page}
		%% \fntext[label2]{}
		%% \cortext[cor1]{}
		%% \affiliation{organization={},
			%%             addressline={},
			%%             city={},
			%%             postcode={},
			%%             state={},
			%%             country={}}
		%% \fntext[label3]{}
		
		\title{Space-time observation of the dynamics of soliton collisions in a recirculating optical fiber loop}
		
		%% use optional labels to link authors explicitly to addresses:
		%% \author[label1,label2]{}
		%% \affiliation[label1]{organization={},
			%%             addressline={},
			%%             city={},
			%%             postcode={},
			%%             state={},
			%%             country={}}
		%%
		%% \affiliation[label2]{organization={},
			%%             addressline={},
			%%             city={},
			%%             postcode={},
			%%             state={},
			%%             country={}}
		
		\author[1]{Fran\c{c}ois Copie}
		\author[1]{Pierre Suret}
		\author[1]{St\'ephane Randoux}
		
		\affiliation[1]{Univ. Lille, CNRS, UMR 8523 - PhLAM - Physique des Lasers Atomes et Molecules, F-59000 Lille, France}

		\begin{abstract}
			We present experiments performed in a recirculating fiber loop in which we realize the single-shot observation of the space and time interaction of two and three bright solitons. The space-time evolutions observed in experiments provide clear evidence of a nearly-integrable nonlinear wave dynamics that can be easily interpreted within the framework of the inverse scattering transform (IST) method. In particular collisions between solitons are found to be almost perfectly elastic in the sense that they occur without velocity change and with only a position (time) shift quantitatively well described by numerical simulations of the integrable nonlinear Schrödinger equation. Additionally our experiments provide the evidence that the position (time) shifts arising from the interaction among three solitons are determined by elementary pairwise interactions, as it is well known in the IST theory. 
		\end{abstract}

		
	\end{frontmatter}
	
	%% \linenumbers
	
	%% main text
	\section{Introduction}\label{sec:intro}
	
	The conditions under which an intense electromagnetic beam can produce its own dielectric waveguide and propagate without spreading were examined by Chiao {\it et al.} as early as 1964 in Ref. \cite{chiao_self-trapping_1964}. In 1973, Hasegawa and Tappert pointed out that nonlinearity of the index of refraction could indeed compensate the pulse broadening effect of chromatic dispersion in low-loss optical fibers leading to undistorted propagation of short pulses. In particular, they theoretically and numerically investigated propagation of light pulses using the one-dimensional nonlinear Schr\"odinger equation (1D-NLSE) both in normal and anomalous dispersion regimes \cite{hasegawa_transmission_1973-1, hasegawa_transmission_1973}. Following this, Mollenaueur {\it et al.} experimentally reported for the first time in 1980 the propagation of solitons in a single-mode optical fiber \cite{mollenauer_experimental_1980}.
	
	The discovery of the optical soliton immediately led to many attempts at applications in ultra-high-speed fiber communications, see Ref. \cite{hasegawa_optical_2022} for a recent historical review on this subject. Many works have been realised to construct and to study all-optical transmission systems in which fiber losses are compensated by optical amplification, see e.g. Ref.  \cite{hasegawa_solitons_1995, taylor_optical_2005, ablowitz_optical_2000, agrawal_fiber-optic_2022, haus_solitons_1996} for relevant monographs and reviews that are available on this subject. In particular Mollenauer and Smith carried out pioneering experiments on long distance optical communications using solitons \cite{mollenauer_demonstration_1988}. They developed a fiber loop system into which a bit stream is loaded and recirculates, thereby simulating propagation over very large distances using a few hundred kilometers of fiber. Recirculating fiber loops are systems where the concept of all optical soliton transmission by periodic amplifications has been investigated in details \cite{mollenauer_soliton_1986} and where soliton propagation has been achieved over distances as large as one million kilometers \cite{nakazawa_10_1991}. 
	
	From a physical point of view, solitons can be described as nonlinear wavepackets that propagate without changing shape, thanks to a balance between nonlinearity and dispersion of the wavepacket. From a mathematical point of view, solitons represent specific  solutions of some nonlinear dispersive partial differential equations (PDEs), such as the 1D-NLSE, which are integrable and can be solved using the inverse scattering transform (IST) method. In the IST theory of the 1D-NLSE, a fundamental soliton is parameterised by its amplitude and velocity, each of these parameters being encoded into a single complex discrete eigenvalue $\lambda_1$ of the associated Zakharov-Shabat scattering problem \cite{zakharov_exact_1972, novikov_theory_1984, yang_nonlinear_2010, taylor_optical_2005}. Due to the integrabillity of the 1D-NLSE, the discrete eigenvalue $\lambda_1$ is a constant of the evolutionary motion, which means that it is preserved with the propagation distance in an optical fiber experiment. A fundamental soliton in the 1D-NLSE is also characterized by its position in space and by its phase. In the IST theory, these phase and position parameters, which generally change with the evolution variable, are encoded in a complex coefficient $C_1$ termed norming constant \cite{zakharov_exact_1972, gelash_strongly_2018}. Noteworthy, the concepts of the IST theory are now used in optical fiber communication where the nonlinear Fourier transform (NFT), originally introduced as eigenvalue communication \cite{hasegawa_eigenvalue_1993}, is now used for optical data processing and fiber transmission, see. e.g. Ref. \cite{turitsyn_nonlinear_2017}. 
	
	As originally noted in the pioneering work by Zabusky and Kruskhal, solitons in nonlinear wave systems described by integrable PDEs exhibit the remarkable property that they retain their shape, amplitude and velocity upon interactions with other solitons \cite{zabusky_interaction_1965}. The process of elastic collision between two solitons occurs without energy exchange between them. It has been studied experimentally in great detail in many physical systems \cite{ikezi_formation_1970, nguyen_collisions_2014, aossey_properties_1992, slunyaev_laboratory_2017, mitschke_experimental_1987, andrekson_observation_1990, aitchison_experimental_1991, shalaby_experimental_1992, parker_collisions_2008, stellmer_collisions_2008, xin_evidence_2021, xin_intense_2022}. In particular, the elastic collision between two individual solitons parametrised by discrete eigenvalues $\lambda_1$ and $\lambda_2$ is accompanied by phase/position shifts that can be measured far outside the interaction region for each soliton. As it is well known in the IST theory, a soliton with the spectral parameter $\lambda_1$ that interacts with another soliton with the spectral parameter $\lambda_2$ experiences a position shift $\Delta_{1 \leftrightarrow 2}$ that only depends on $\lambda_1$ and $\lambda_2$: $\Delta_{1 \leftrightarrow 2}=\Delta(\lambda_1,\lambda_2)$, and not on the position and phase parameters that are encoded in the norming constants $C_1$ and $C_2$ associated with each soliton \cite{novikov_theory_1984, yang_nonlinear_2010, taylor_optical_2005}. In an interaction process involving three solitons with spectral (IST) parameters $\lambda_j$ $(j=1,2,3)$, the IST theory predicts that the total position shift $\Delta_{1 \leftrightarrow (2,3)}$ experienced by the soliton with the spectral parameter $\lambda_1$ is equal to the algebraic sum of the shifts accumulated during the paired collisions: $\Delta_{1 \leftrightarrow (2,3)}=\Delta_{1 \leftrightarrow 2}+\Delta_{1 \leftrightarrow 3}$ \cite{zakharov_exact_1972}. This fundamental property that solitons exhibit particle-like properties with elastic, pairwise interactions arises from the integrable nature of the PDE describing their evolution in space and time.
	
	Beyond regular solitons that are localised structures sitting on zero background, the 1D-NLSE supports a broad class of solutions referred to as breathers or solitons on finite background which includes the so-called Akhmediev, Peregrine and Kuznetsov-Ma breathers \cite{dudley_instabilities_2014}. The interaction of such breather waves features an even richer dynamics that has been investigated experimentally in several domains including fiber optics and hydrodynamics systems \cite{kibler_superregular_2015, xu_ghost_2020, gelash_management_2022}.

	In this paper, we present experiments performed in a recirculating fiber loop in which we observe the interaction in space and time of some small ensembles composed of two or three bright solitons. Despite the presence of weak dissipation occurring over propagation, the observed dynamics is found to be nearly integrable, which means that soliton collisions are found to be nearly elastic without significant changes of their velocities due to the interactions with other solitons. Considering the collision between two bright solitons, we observe that their relative phase determines the shape of the field in the interaction region where their overlap is significant. On the other hand, the position (time) shift measured far outside the interaction region only depends on the amplitudes and velocities of the two solitons without being influenced by the phase-dependent dynamics of the field in the interaction region. Considering the interaction between three bright solitons, we observe that the position shift depends on the spectral IST parameters (amplitude and velocity) of the solitons without depending on their position and phase parameters. The space-time behaviours observed in the experiments are reproduced quantitatively well by numerical simulations of the 1D-NLSE. The space-time evolutions recorded in the experiments provide the clear evidence of almost perfectly integrable nonlinear wave dynamics that can be easily interpreted within the framework of the IST theory.

	\section{Experimental setup}\label{sec:setup}
	
	\begin{figure*}[th!]\centering
		\includegraphics[width=.85\textwidth]{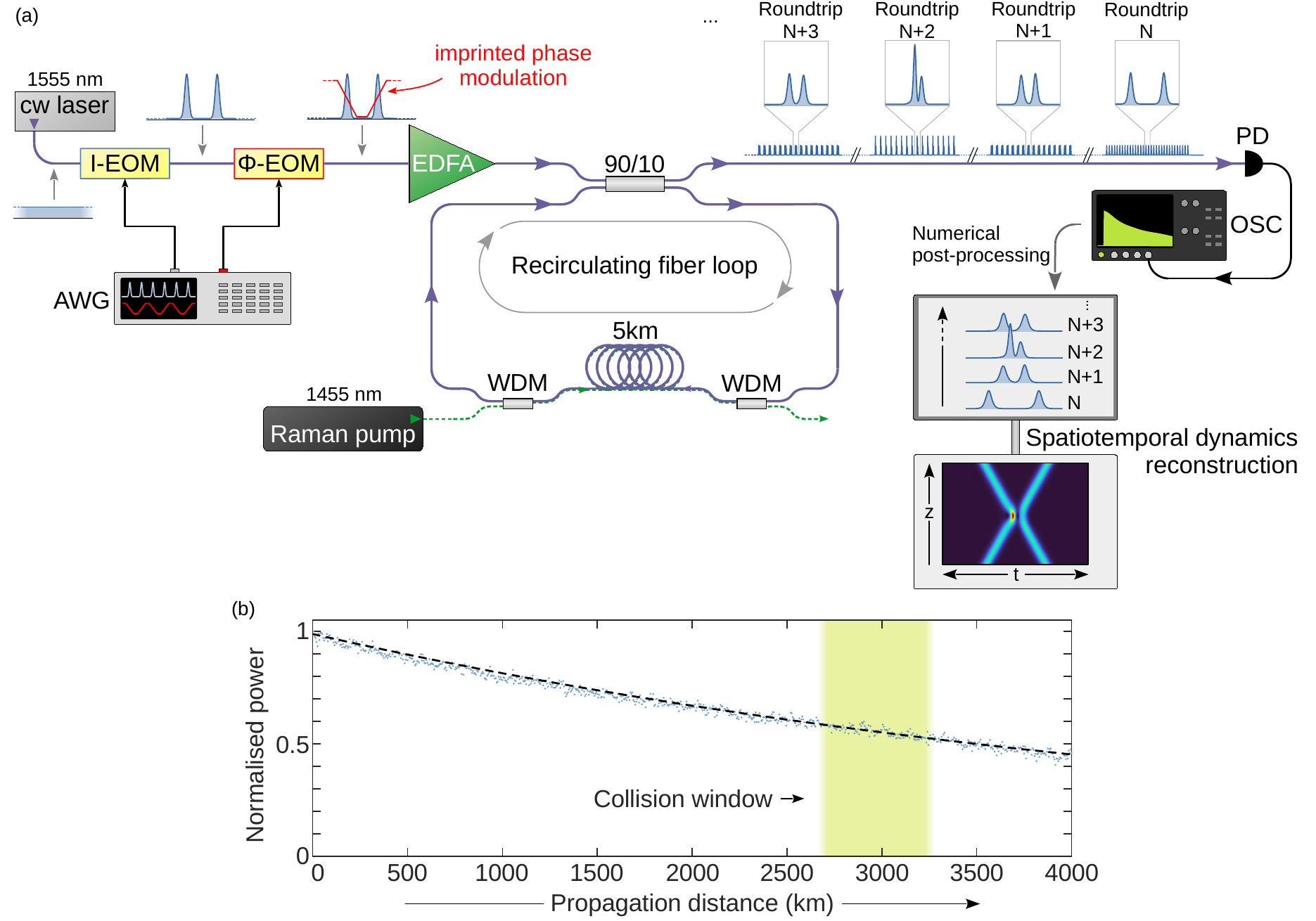} 
		\caption{(a) Principle of operation of the experimental setup. (b) Normalised evolution of the circulating optical power computed from the experimental recording from which data presented in Fig.\,\ref{fig:2sol_phase} are extracted. Light blue dots are experimental values calculated at each roundtrip and the black dashed line is the exponential fit  giving $\alpha_{\text{eff}} \sim \SI{1.92e-4}{\per \kilo\meter}$. cw: Continuous Wave, I-EOM: Intensity-Electro-Optic Modulator, $\phi$-EOM:  Phase-Electro-Optic Modulator, EDFA: Erbium Doped Fiber Amplifier, WDM: Wavelength Division Multiplexer, PD: Photodetector, OSC: Oscilloscope.
			\label{fig:setup}
		}
	\end{figure*}

	The principle of operation of the experimental setup is schematically depicted in Fig.\,\ref{fig:setup}(a). It is essentially similar to the one reported in Ref. \cite{kraych_nonlinear_2019, kraych_statistical_2019, copie_spatiotemporal_2022} and used for experimental investigations of the nonlinear stage of modulation instability of a plane wave in various configurations. A continuous wave field from a narrow-band laser at \SI{1555}{nm} is temporally shaped using a combination of \SI{20}{GHz} bandwidth electro-optic (intensity) modulator (I-EOM) and phase modulator ($\phi$-EOM). A \SI{12.5}{GHz} arbitrary waveform generator (AWG) drives the I-EOM to generate a train of short pulses and synchronously drives the $\phi$-EOM to imprint a linear phase across each pulse with an alternating sign of the slope. Using this synchronous phase modulation scheme, the group velocity of each pulse can be controlled in an accurate way.
	
	In the experiments reported in the present paper, the phase modulation is designed in such a way that adjacent pulses have opposite velocities in order to trigger their collision at the desired propagation distance. After the modulation stage, the wave field is amplified to reach Watt-level peak power and chopped down by an acousto-optic modulator (not shown for clarity) to create a \SI{1}{\micro s}-long signal burst at a \SI{20}{Hz} repetition rate which is used as input of the recirculating optical fiber loop.
	
	The fiber loop itself is mostly composed of $\sim$ \SI{5}{km} of single mode fiber (SMF) closed on itself by a $90/10$ fiber coupler. The coupler is arranged in such a way that $10\,\%$ of the signal is initially injected in the loop and $90\,\%$ of the intra-loop power is recirculating. The optical signal circulates in the clockwise direction and at each roundtrip, $10\,\%$ of the circulating power is extracted and directed towards a photodetector (PD) coupled to a fast sampling oscilloscope leading to an overall \SI{32}{GHz} detection bandwidth. Experimental data are recorded with the oscilloscope at a sampling rate of $160$ GSa/s. They consist in a succession of sequences (one per roundtrip) that are subsequently processed numerically to construct single shot space-time diagrams showing the wavefield dynamics. Note that in addition to the short pulses, the input signal contains flat-top pulses of durations ranging from \SI{500}{ps} to \SI{10}{ns} (not illustrated in Fig.\,\ref{fig:setup}) that are used as reference markers for post-processing synchronisation and power calibration. Importantly, the losses accumulated over one circulation in the fiber loop are partially compensated using a counter-propagating Raman pump coupled in and out of the loop via wavelength division multiplexers (WDMs). As illustrated in Fig.\,1(b) which shows the normalised evolution of the power in a typical experimental run, this enables reduction of the effective exponential power decay rate of the circulating field to $\alpha_{\text{eff}} \sim \SI{1.92e-4}{\per \kilo\meter}$ or equivalently $\sim \SI{0.00084}{dB \per km}$. This value can be normalised with respect to the typical nonlinear length of the experiments presented in this work which is $L_{NL} = 1/(\gamma P_0) = \SI{14.7}{km}$ (see caption of Fig.\,\ref{fig:2sol_phase} for the values of the parameters). Accordingly, the losses expressed in unit of nonlinear length is $\alpha_{\text{eff}} L_{NL} = 2.8 \cdot 10^{-3}$.

	As already mentioned in Sec.\,\ref{sec:intro}, recirculating fiber loops represent useful platforms for the research and development of long-haul transmission systems. In configurations typically implemented in the context of optical communications, the fiber loop system is usually first loaded with a bit stream which subsequently circulates over a given number of roundtrips before the detection/diagnostic of the optical field is made. The architecture of such loop systems is based on the use of several optical switches that must be appropriately activated at the loading, circulating and detection stages, see e.g. Ref. \cite{bergano_circulating_1995}. Such a fiber loop device has been recently used for the space-time observation of Fermi-Pasta-Ulam-Tsingou recurrence in a long-haul optical fiber transmission system \cite{goossens_experimental_2019}. In contrast with these loops incorporating optical switches, our system is based on a fiber coupler which is used both to inject and to extract light, thereby using an architecture similar to that proposed in Ref. \cite{desurvire_raman_1985}. The recirculating fiber loop shown in Fig.\,1(a) has the strong advantage of allowing real-time periodic recording of the evolution of the temporal dynamics of a signal at each roundtrip. In other words, our experimental setup permits the reconstruction of the spatiotemporal dynamics of a signal in single shot, which enables the observation of the dynamics of non-repetitive events. 
	
	\section{Collision of two solitons}\label{sec:two_solitons}

\begin{figure*}[!t]
		\includegraphics[width=1\textwidth]{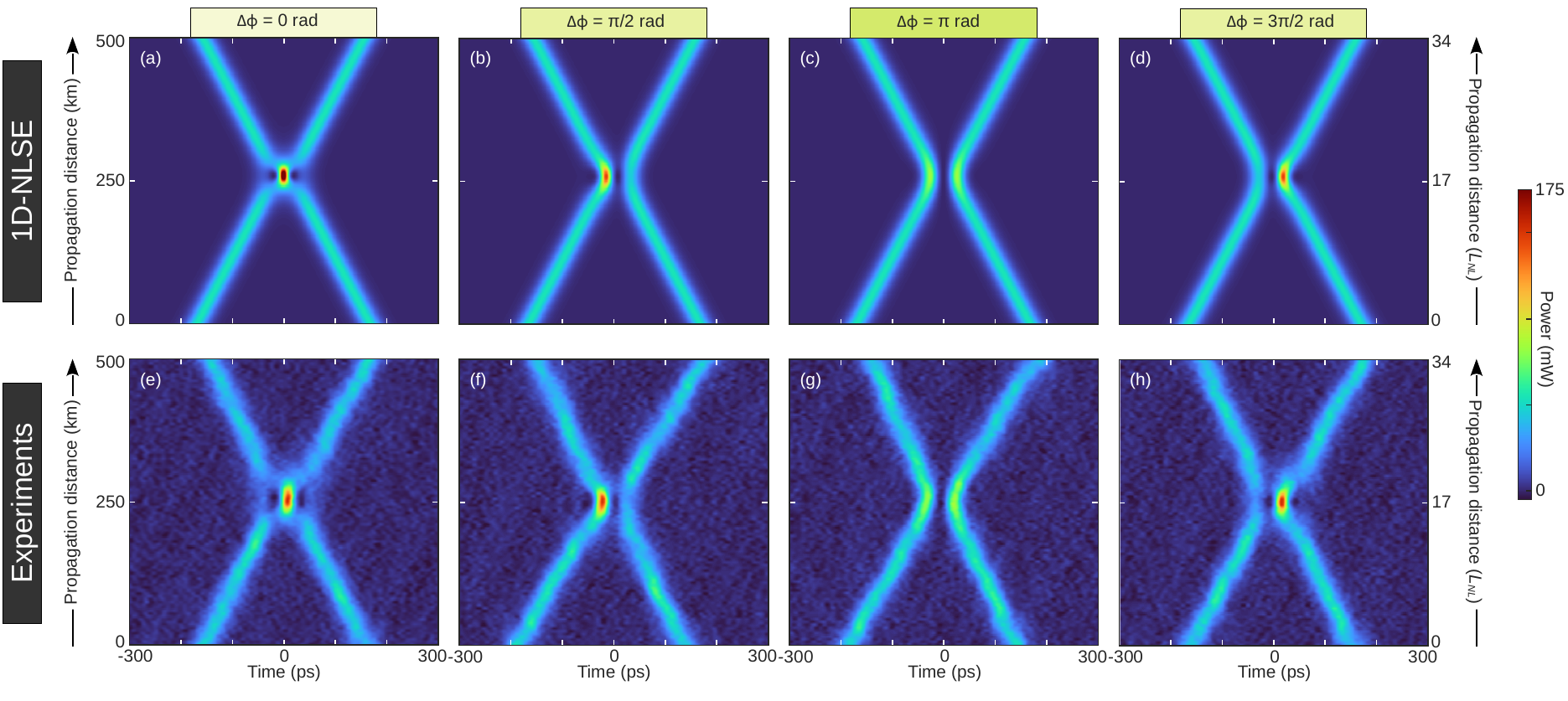}
		\caption{Spatiotemporal dynamics of the collision of two identical solitons, impact of the phase difference. (a-d) Numerical simulations of the 1D-NLSE for 4 remarkable relative phase differences. (e-h) Corresponding experimental observations selected from a batch of 50 collisions. Vertical axis is shifted for direct comparison since collisions actually occur after $\sim \SI{3000}{km}$ of propagation. The color scale is common to all the presented results. Initial condition for the simulations is of the form $\Psi(t, z = 0) = \sqrt{P_0}\text{sech}[(t - t_0)/T_0]e^{-i(\Delta\omega/2) t} + \sqrt{P_0}\text{sech}[(t + t_0)/T_0]e^{+i((\Delta\omega/2) t + \Delta\phi)}$, with $P_0 = \SI{55.3}{mW}, \beta_2 = \SI{-22}{ps \squared/km}, \gamma = \SI{1.23}{\per \watt \per \kilo\meter}, T_0 = \sqrt{|\beta_2|/(\gamma P_0)} \sim \SI{18}{ps}, \Delta \omega/(2\pi) = \SI{8.62}{GHz}, t_0 = \SI{168}{ps}$.
			\label{fig:2sol_phase}
		}
	\end{figure*}

	Using the experimental system described previously, we control with a satisfying accuracy the process of collision between individual solitons and observe the associated spatiotemporal dynamics in single shot. Firstly, we designed the pulses launched in the fiber loop to have a typical duration of $35.7 \pm \SI{0.9}{ps}$ FWHM and a peak power of $101 \pm \SI{4}{mW}$ in such a way that their initial shape is compatible with fundamental solitons. However, the soliton generation process is not perfect and our initial pulses unavoidably contain some radiative (non-solitonic) content. Importantly, owing to the slight variations of the pulses parameters from shot to shot, the phase difference between adjacent pulses is not controlled in the experiment. The synchronous phase modulation imprints a linear phase slope to each pulse which results in a relative velocity difference between adjacent pulses that can be translated in relative linear drifts $\Delta V \sim 1.2 \pm \SI{0.04}{ps \per km}$ or equivalently a frequency detuning $\Delta f = \Delta \omega/(2\pi) = \Delta V /(2\pi\,|\beta_2|) \sim 8.62 \pm \SI{0.3}{GHz}$. This is calculated a posteriori by measuring and fitting the trajectories of the solitons in the reconstructed spatiotemporal diagrams. A single experimental recording features several tens of soliton collisions from which the fluctuations given above are estimated. The input signal is engineered such that collisions occur after a propagation distance of $\sim \SI{3000}{km}$ which allows the initial radiative content of the pulses to be cleared out (see the collision window in Fig.\,\ref{fig:setup}(b)). That way, we ensure that the collisions involve almost purely solitonic pulses with a limited contribution from dispersive waves. The characteristics of the temporal shape of the pulses are calculated via fitting with $\text{sech}^2$ functions (see caption of Fig.\ref{fig:2sol_phase} for the expression) \SI{250}{km} before collision and are as follows: $P_0 = 55.3 \pm \SI{8}{mW}$, $T_0 = 19.9 \pm \SI{3}{ps}$ which gives $N = \sqrt{\gamma P_0 T_0^2 /|\beta_2|} = 1.11 \pm 0.1$ \cite{agrawal_nonlinear_2013}.
	
	Figure \ref{fig:2sol_phase}(e-h) shows the spatiotemporal dynamics of four different collisions recorded experimentally within a single experimental run including a total of 50 collision events. Each collision event is presented over a propagation distance range of \SI{500}{km} with the collision occurring after a propagation distance that has been artificially shifted for simplicity around \SI{250}{km} (the actual collision occur after a propagation distance of $\sim \SI{3000}{km}$ inside the loop). It is noteworthy that this propagation distance corresponds to 34 nonlinear lengths ($L_{NL} = \SI{14.7}{km}$). The space-time diagrams in  Fig.\,\ref{fig:2sol_phase} are presented in a reference temporal frame moving at a group velocity corresponding to the mean group velocity of the two pulses, which results in collisions that appear nearly symmetric with respect to the vertical axis ($t = 0$). The relative phase between colliding solitons not being controlled in the generation process, the interaction region takes a different shape depending on this phase that is randomly distributed between $0$ and $2\pi$.
	
	We show in Fig.\,\ref{fig:2sol_phase}(a-d) numerical simulations of the 1D-NLSE expressed with physical variables
	
	\begin{equation}
		i\frac{\partial \Psi}{\partial z} = \frac{\beta_2}{2}\frac{\partial^2\Psi}{\partial t^2} - \gamma|\Psi|^2\Psi,
	\end{equation}
	where $\Psi$ is the complex envelope of the wavefield, $\beta_2$ and $\gamma$ the group velocity dispersion and Kerr coefficients respectively, $z$ the longitudinal variable describing propagation distance, and $t$ the time in the reference frame introduced previously. We take as initial conditions two well separated solitons
	
	\begin{equation}
		\begin{split}
			\Psi(t, z = 0) = \sqrt{P_0}\text{sech}[(t - t_0)/T_0]e^{-i(\Delta\omega/2) t}\\
			+ \sqrt{P_0}\text{sech}[(t + t_0)/T_0]e^{+i[(\Delta\omega/2) t + \Delta\phi]},
		\end{split}
	\end{equation}
	with peak power $P_0$ and frequency detuning $\Delta \omega$ fixed to the average value experimentally estimated before collision, their typical duration $T_0$ is set to the exact value leading to fundamental solitons (see parameters' values in the caption of Fig.\,\ref{fig:2sol_phase}). Spatiotemporal evolution for four particular values of $\Delta \Phi$ are illustrated: in-phase ($\Delta\Phi = 0$) collision (a) is translated into a strong peak of the optical intensity whereas the pattern resulting from out-of-phase ($\Delta\Phi = \pi$) collision (c) creates the false impression that solitons exactly bounce (or repel) each other while they still actually pass through one another. Intermediate phase differences $\Delta\Phi = \pi/2$ and $3\pi/2$ respectively lead to similar evolutions where a peak optical intensity appears but shifted towards the left (right) with respect to the collision center.
	
	Experimental observations presented in Fig.\,\ref{fig:2sol_phase}(e-h) have been respectively selected for their striking quantitative agreement with the simulations illustrated above. Each of the experimental realisations, not shown here for brevity, matches with numerical simulation with appropriate phase difference. This confirms that we indeed realise the collision of solitonic pulses, the coherence of which is well preserved all along their interaction. Note that numerical simulations do not take account of dissipation that is present in experiment but we stress that the counter-propagating Raman amplification limits the power loss over the \SI{500}{km} shown in Fig. \ref{fig:2sol_phase}(e-h) to less than $10\,\%$ which barely affects the dynamics with respect to the conservative case.
	
	\begin{figure}[!t]\centering
		\includegraphics[width=.4\textwidth]{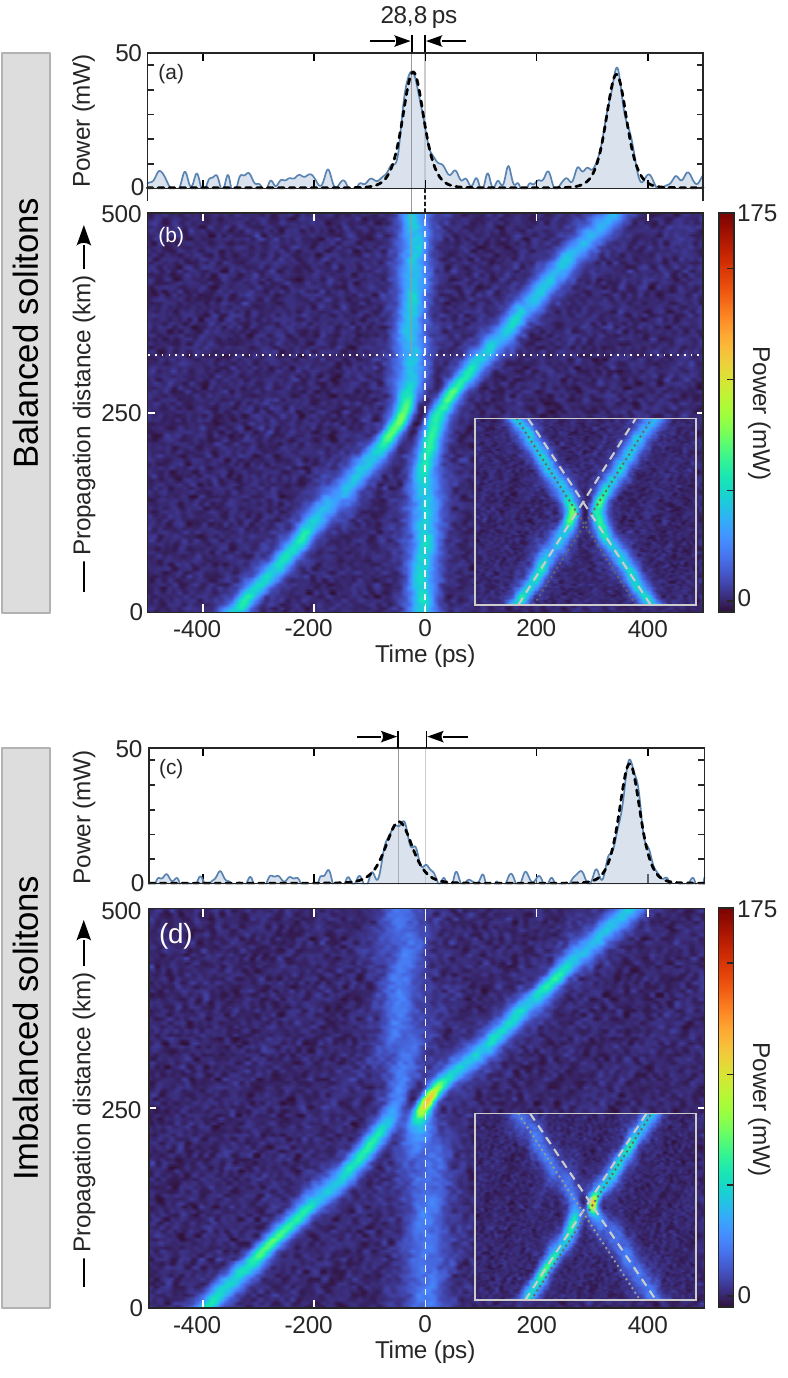}
		\caption{Experimental observation of collision-induced soliton shift in (a-b) balanced and (c-d) imbalanced configurations. The space-time diagrams are plotted in the reference frame of the right soliton for clarity.  (a, c) Temporal traces extracted from the top of the space-time diagrams with $\text{sech}^2$ fits superimposed. Vertical dashed white lines in (b) and (d) highlight the linear trajectories of the soliton before collision. Insets in (b) and (d) show the same collisions in the temporal reference frame that travels at the mean group velocity of the two pulses.
			\label{fig:shift}
		}
	\end{figure}

	As exposed in the introduction, a remarkable feature of solitons is that they emerge unaltered from their pairwise interaction although their relative phase difference drastically impacts the exact dynamics of the collision as confirmed by the experiments presented Fig.\,\ref{fig:2sol_phase}. However, the collision is accompanied by a time shift that is solely parametrised by the IST spectral parameters of the individual solitons (amplitude and velocity).
	
	Our experiments enable clear observation of the collision-induced temporal shift as presented in Fig.\,\ref{fig:shift}. We first take the case of the collision between \textit{balanced} solitons (i.e. solitonic pulses with comparable amplitudes) similar to those illustrated in Fig.\,\ref{fig:2sol_phase} and focus on the temporal shift experienced by one of the soliton. To do so, it is convenient to operate a change of temporal reference frame to one that travels at the group velocity of the soliton under consideration. That way, the latter appears as following a straight vertical trajectory when not interacting with other solitons. This operation is done numerically a posteriori for ease of visualisation and does not modify the physics of the system. Figure \ref{fig:shift}(b) shows the space-time dynamics of a collision in the same conditions as in Fig.\,\ref{fig:2sol_phase} but plotted in the reference frame of the \textit{right} (\textit{steady}) soliton. The vertical dashed white line emphasises the trajectory of the soliton before collision. After collision with the \textit{left} (\textit{moving}) soliton, it has clearly experienced a temporal shift towards negative times of the order of the soliton's width while retaining almost perfectly its velocity close to zero after the interaction, thus demonstrating that the observed collision is of elastic nature.
	
	The temporal trace recorded $\sim \SI{250}{km}$ after the collision is shown in Fig.\,\ref{fig:shift}(a) in which the collision-induced shift is marked by the arrows. The two pulses are well fitted by $\text{sech}^2$ functions (dashed lines) whose parameters still confirm their solitonic nature. Note that the signal exhibit significant noise due to the rather small optical power detected in single shot (we recall that less than $10\,\%$ of the circulating power is detected so the $\sim \SI{50}{mW}$ peak power indicated in Fig.\,\ref{fig:shift}(a) correspond effectively to pulses of less than $\sim \SI{5}{mW}$ peak power incident on the photodiode). In this configuration, the \textit{moving} soliton also experiences a shift of almost the same magnitude but opposite sign (as visible in the inset showing the same collision in the reference frame that travels at the mean group velocity of the two pulses).
	
	As previously mentioned, we have recorded 50 similar realisations of this scenario of collision which allows us to evaluate an average shift of the solitons with a certain error interval. To do so, we extracted from the spatiotemporal diagrams the linear trajectories of the two solitons before and after interaction for each realisation. The collision-induced shift is then calculated from the parameters of the linear trajectories. Note that parameters of each collisions (amplitude, duration of the solitons and their relative velocities) are slightly different such that the expected temporal shift also varies from shot to shot. Also, care must be taken since small changes of solitons' velocity are detected between before and after collision which is a consequence of weak non-integrability of our system. This prevents straightforward estimation of the temporal shift which is formally defined at asymptotically long propagation distance from the interaction \cite{zakharov_exact_1972, novikov_theory_1984}. To limit the impact of this velocity change, we estimated the shift \SI{75}{km} after the point of maximum interaction (marked by the horizontal dotted white line) which is sufficiently large for the solitons to separate but not too much for the change of velocity to dramatically alter the measurement of the shift. Using this methodology, the average temporal shift measured is $29.1 \pm \SI{4.2}{ps}$. For comparison, numerical simulations of the 1D-NLSE taking account of the realistic variation of all the relevant parameters give a shift of $30.7 \pm \SI{0.9}{ps}$. Eventhough our experimental estimation is associated to a larger uncertainty, it remarkably agrees with realistic prediction from the integrable 1D-NLSE model.
	
	To complement the result shown in Fig.\,\ref{fig:shift}(a-b), we present an example of a so-called imbalanced collision, i.e. a collision between solitons of significantly different peak power. The spatiotemporal evolution of the pulses plotted in the reference frame moving at the velocity of the weakest soliton is shown in Fig.\,\ref{fig:shift}(d) and exhibit a similar dynamics as the one described previously. For this particular realisation, the weak soliton clearly experiences a large temporal shift and keeps its velocity, demonstrating once again that the observed imbalanced collision is of elastic nature. The inset in Fig.\,\ref{fig:shift}(d) additionally illustrates that the weak soliton emerges from the collision with a larger shift than the strong one.
	
	\section{Three solitons interaction}\label{sec:three_solitons}
	
	In this last section, we report experiments on the interaction process between three solitons. In this configuration, the IST theory predicts that the total shift experienced by one of the soliton is the algebraic sum of the shifts acquired over the paired collisions. The result of the interaction between three solitons is thus fully determined by the three constitutive pairwise collisions. Importantly, the relative positions and phase of each soliton before the interaction do not influence the time shift that are measured well after the interaction among the three solitons. 
	
	\begin{figure*}[!t]
		\includegraphics[width=1\textwidth]{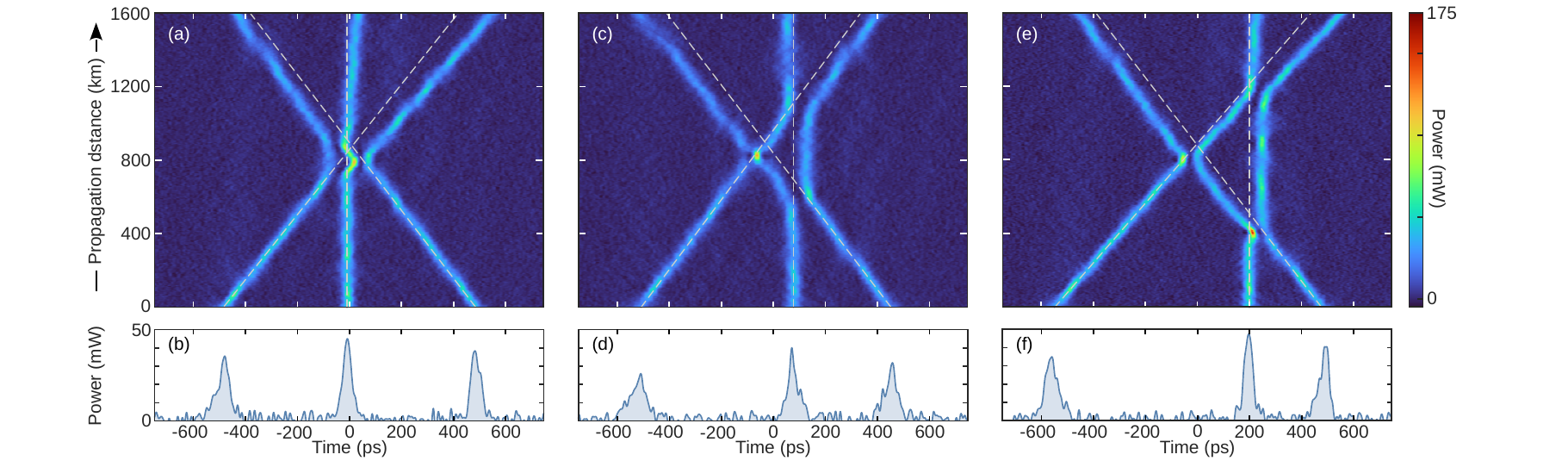}
		\caption{Spatiotemporal dynamics of three solitons interaction for various initial positions of the solitons (from evenly (a) to largely unequally (c) spaced). Dashed white lines highlight the linear trajectories of the solitons before interaction. (b, d, f) Temporal traces at the bottom of the corresponding space-time diagrams (i.e. before interaction). 
			\label{fig:3sol}
		}
	\end{figure*}
	
	To illustrate this, we designed a set of experiments where three solitons of nearly the same amplitude (comparable to that shown in Fig.\,\ref{fig:2sol_phase}) interact and we show the results in Fig.\,\ref{fig:3sol}. The velocity and initial position of the solitons are precisely controlled to generate three distinct interaction scenarios. In all cases, frequency detuning (relative velocity differences) between adjacent pulses is constant and equal to $\Delta f \sim \SI{4.24}{GHz}$ ($\Delta V \sim \SI{0.59}{ps \per km}$). Also, because the total interaction between the three solitons takes place over a larger span of distance we engineered the input signal so that it occurs at shorter propagation distance compared to the previous experiments (centered around \SI{1000}{km} of propagation instead of \SI{3000}{km}) which provides better control over the solitons parameters. This means that a greater dispersive content is present in the interaction area with respect to previous experiments. The dynamics of the solitons is therefore visibly impacted by this interplay.
	
	In the first case illustrated in Fig.\,\ref{fig:3sol}(a), the solitons are initially equally spaced leading to an interaction that is very localised in space and time. In fact, individual pairwise collisions are not distinguishable and the exact collision dynamics depends on the relative phase differences between the three solitons. This results in a greater complexity of the interaction region as compared to the dynamics associated with a two soliton interaction pictured in Fig.\,\ref{fig:2sol_phase}. Noticeably, the central soliton emerges almost unshifted from the interaction since in this configuration, the shifts induced from the two indivual collisions is in principle of the same magnitude but of opposite sign. The two other solitons experience a total shift larger than those observed in Fig.\,\ref{fig:shift} because they both cumulate two shifts of same sign. Also, it is clear that the velocity of the central soliton has slightly changed during the interaction. The weak but non null effective losses and the periodic evolution of the power imposed by the recirculating loop configuration break the integrability of the system to some extent and are involved in the slightly inelastic features observed in the experiments.
	
	If the in-between soliton is now initially slightly offsetted, the interaction window expands and pairwise separable collisions start to become apparent though individual trajectories still cannot be identified as shown in Fig.\,\ref{fig:3sol}(c). Again, the dynamics of the interaction is highly sensitive to the relative phases between the solitons. However, the central soliton comes out of the interaction quasi unperturbed in terms of position and velocity. Finally, for a large offset of the in-between soliton the three individual pairwise collisions are perfectly identified (see Fig.\,\ref{fig:3sol}(e)) and each exhibit a dynamics comparable to those reported in Fig.\,\ref{fig:2sol_phase} and \ref{fig:shift}. Paying specific attention to the in-between soliton, the latter experiences a clear positive shift from the first collision that is almost perfectly cancelled after the second collision. This is overall remarkably translated by its broken but straight trajectory. The experimental observations presented in Fig.\,\ref{fig:3sol} illustrate qualitatively yet with a great level of details the full complexity of the particle-like behaviour of solitons' pairwise interaction. 
	
	\section{Conclusion}\label{sec:conclusion}
	
	In conclusion, a recirculating fiber loop has been used to realize the single-shot observation of the interaction in space and time of two and three bright solitons. To the best of our knowledge, it is the first time in nonlinear fiber optics that the fundamental process of soliton collision is observed both in single-shot (without averaging) and in space and time. In our experiments, collisions between two solitons have been found to be almost perfectly elastic, without significant change in the velocity of the solitons after their interaction. The position (time) shift measured in experiments is quantitatively well described by numerical simulations of the 1D-NLSE. Our experiments have also provided the evidence that the position (time) shifts arising from the interaction among three solitons are determined by the elementary pairwise interactions, as it is well known in the IST theory. 
	
	The recirculating fiber loop is a weakly dissipative experimental platform that is well adapted to the space-time observation of the evolution of solitons and solitary waves. We plan to use it in the near future to investigate theoretical questions behind the spectral theory of soliton gases \cite{el_spectral_2020, suret_soliton_2023} and generalized hydrodynamics, the hydrodynamic theory of many-body quantum and classical integrable systems \cite{doyon_lecture_2020}.
	
	\section*{Acknowledgment}
	The authors are very grateful to H. Damart, and G. Dekyndt for the technical support and to A. Amo for providing access to the fast waveform generator.
	This work has been partially supported by the Agence Nationale de la Recherche through the StormWave (ANR-21-CE30-0009) and SOGOOD (ANR-21-CE30-0061) projects, the LABEX CEMPI project (ANR-11-LABX-0007), the Ministry of Higher Education and Research, Hauts de France council and European Regional Development Fund (ERDF) through the Contrat de Projets Etat-Région (CPER Photonics for Society P4S). The authors would like to thank the Isaac Newton Institute for Mathematical Sciences for support and hospitality during the programme “Dispersive hydrodynamics: mathematics, simulation and experiments, with applications in nonlinear waves” when part of the work on this paper was undertaken.
	
	%% The Appendices part is started with the command \appendix;
	%% appendix sections are then done as normal sections
	%% \appendix
	
	%% \section{}
	%% \label{}
	
	%% If you have bibdatabase file and want bibtex to generate the
	%% bibitems, please use
	%%
	%%  \bibliographystyle{elsarticle-num} 
	%%  \bibliography{<your bibdatabase>}
	
	%% else use the following coding to input the bibitems directly in the
	%% TeX file.
	
	\bibliographystyle{elsarticle-num} 
	\bibliography{Article_SolCol}
	
	%\begin{thebibliography}{00}
	
	%% \bibitem{label}
	%% Text of bibliographic item
	
	%\bibitem{}
	
	%\end{thebibliography}
\end{document}